# Performance Evaluation of Ad Hoc Multicast Routing Protocols to Facilitate Video Streaming in VANETS

Muhammad Danish Khan[1], Arshad Shaikh[2], Hameedullah Kazi[3]

[1]Isra University Hyderabad, PAKISTAN
[2]Department of Computer Science, Isra University Hyderabad, PAKISTAN
[3]Department of Compute Science, Isra University Hyderabad, PAKISTAN

Vehicular Ad Hoc Network (VANET) is a type of mobile ad hoc network (MANET) that facilitates communication among vehicles. VANET provides inter-vehicular communications to serve for the application like road traffic safety and traffic efficiency. Infotainment service has been an anticipating trend in VANETs, and video streaming has a high potential in VANET. Although, this emerging technology is trending, there are still some issues like QoS provisions, decentralized medium access control, node coverage area, and finding and maintaining routes due to highly dynamic topology. These issues make multicast communication difficult in VANETs. Numerous routing protocols and routing strategies have been projected to cope with these issues. Lots of work has taken place to assess and measure the performances of these protocols in VANETs but these protocols are rarely analyzed for performance under stress of real time video multicast.

In this study two different multicast routing protocols viz. Multicast Ad hoc On Demand Distance Vector (MAODV) and Protocol for Unified Multicasting through Announcements (PUMA) are evaluated for facilitating video streaming in VANETS. The protocols are examined against the QoS parameters such as Network Throughput, Packet Delivery Ratio (PDR), Average end to end Delay, and Normalized Routing Load (NRL). Variable Bit Rate (VBR) traffic is used to evaluate the performances of protocol. PUMA, at the end, showed better performance against different QoS provisions in different scenarios.

## I. INTRODUCTION

Recent advancements in communication are enabling the design and implementation of networks those can be deployed in various kinds of environments, this network is called an Ad Hoc Network. An infrastructure-less network comprising different number of communication nodes describes an ad hoc network, where a node may be any communication device [12]. Sort of these networks are very useful where traditional wired networks cannot be deployed, such as battlefields, disastrous areas etc. Mobile Ad Hoc Network (MANET) is one such ad hoc network with no Base Station or Access Point available for passing messages among nodes, the nodes are mobile and wirelessly communicate to each other without banking on pre-established infrastructure [19]. Participating node in a MANET might be any equipment with wireless interface embodied in it. Every participating node may also work as router for passing the messages between distant nodes. Because nodes are mobile, there's no bound on the movement of nodes, and each node changes its links frequently.

Specifically when talking about VANET, Vehicular Ad Hoc Network, it is also a MANET where moving vehicles act as nodes to create a mobile network [49]. The participating vehicles are turned either into router or host resulting in a wider network as the small distant vehicles join each other [9]. VANETs have received a lot of interest in the applications like road traffic safety and traffic efficiency [15]. Because there are vehicles working as nodes VANETs have high mobility among the nodes causing highly dynamic topology. Furthermore in the absence of basic infrastructure each node is responsible for finding the path for its data to be transmitted. The routing protocol is set of rules for establishment and maintenance of paths among mobile nodes. VANET in its nature is of highly dynamic topology; nodes join and leave the network frequently causing periodic connectivity information requirement in order to have a reliable status of the network. Consequently, the aim of a routing protocol is to discover right path among mobile nodes so that data can be delivered reliably. Furthermore, the established path for communication should be the one with minimum overhead and bandwidth consumption. During previous decades different routing protocols are projected for VANETs [29].

In this study we have evaluated the performances of two prominent ad hoc multicast protocols viz. Multicast Ad hoc On-Demand Distance Vector (MAODV) a tree based protocol, and Protocol for Unified Multicasting through Announcements (PUMA) that is mesh based. The protocols will be evaluated under stress of video traffic (VBR) in VANETs with multicast mode of communication.

## II. RELATED WORK

### A. A Comparative Study of Tree based Vs. Mesh based Multicast Routing Protocols in Mobile Ad hoc Networks

Reference [7] compared the performance of two routing protocols that do multicasting on the basis of tree structure and mesh structure. The protocols are MAODV and PUMA, tree based and mesh based respectively. Network Simulator (ns-2.35) was used as simulation tool. They have carried out a number of scenarios for examining the performance against nodes mobility impact, varying count of senders, varying number of nodes, and so. The performance measures used were Throughput, End-to-End Delay, and the Packet Delivery Fraction (PDF). Random Waypoint mobility model was adopted for this work. Following table shows the simulation results for either of protocols. Table – I shows the summary of their results.

**Table – I: MAODV vs PUMA**

| Number of Nodes | Routing Protocol | Throughput | End-2-End Delay | PDF |
|---|---|---|---|---|
| 25 | MAODV | 140.91 | 0.03 | 0.6028 |
|  | PUMA | 170.27 | 0.01 | 0.9952 |
| 50 | MAODV | 159.98 | 0.06 | 0.6912 |
|  | PUMA | 728.52 | 0.996 | 0.9895 |
| 75 | MAODV | 172.12 | 0.08 | 0.6601 |
|  | PUMA | 1904.92 | 1.256 | 0.9787 |
| 100 | MAODV | 179.26 | 0.091 | 0.6345 |
|  | PUMA | 4430.70 | 1.2366 | 0.9753 |

*B. Analysis of Multicast Routing Protocols: PUMA and ODMRP*

In their work [45] have evaluated and compared the performance of PUMA and ODMRP routing protocols that do multicasting. Both the protocols are mesh-based on their packet distribution approach. The protocols are reactive protocols. The quantitative measures used as performance metrics are Packet Delivery Ratio (PDR), Average End to End Delay (Average EED), Throughput, and the Normalized Routing Load (NRL). Mathematically,

PDR= (Delivered Packets)/(Sent Packets)
Average EED= (Total EED)/(No of Sent Packets)
Throughput= (Number of Sent Packets)/(Time Taken)
NRL=(Number of Data Packets Recieved)/(Number of Routing Packets Received)

The network simulator (NS2) and Qualnet environment were used to simulate PUMA and ODMRP respectively using Random Waypoint as the mobility model. Table -2 below elaborates the simulation setup. The results have proven less end-to-end delay in PUMA, compared to ODMRP. PUMA also outperformed ODMRP with respect to packet delivery ratio, average end-to-end delay and throughput. The reason behind that is elaborated as "each source do flooding in ODMRP causing a large count of senders and congestion, therefore leading to large packet drop ratio whereas PUMA has only the core node that floods the network". An increase in count of multicast groups leaded to more congestion.

**Table – II: Simulation Parameters for PUMA and ODMRP**

| Simulator | Network Simulator (NS-2) | QualNet 5.0.2 |
|---|---|---|
| Total Nodes | 50 | 50 |
| Simulation Time | 200 sec | 200 sec |
| Simulation Area | 1500x300 | 1500x300 |
| Propagation Model | Two Ray Ground Model | Two Ray Ground Model |
| Pause Time | 0 - 10 sec | 0 -10 sec |
| Mobility Model | Random Waypoint Model | Random Waypoint Model |
| Radio Range | 250 m | 250 m |
| MAC Protocol | MAC_802.11 | MAC_802.11 |
| Data Packet Size | 512 bytes | 512 bytes |
| Data Rate | 11 Mbps | 11 Mbps |
| Antenna | Omni Directional Antenna | Omni Directional Antenna |
| IFQ Length | 50 packets | 50 packets |
| Bandwidth of Physical Layer | 11 Mbps | 11 Mbps |
| No. of Receivers | 10, 20, 30, 40 | 10, 20, 30, 40 |
| Routing Protocol | PUMA | ODMRP |
| Destination Address | 224.0.0.1 | 225.0.0.1 |
| No. of Packets Per Sec | 10 | 10 |
| Mobility Speed | 0 – 10 m/sec | 0 – 10 m/sec |
| Traffic | CBR | MCBR |

*C. PERFORMANCE ANALYSIS OF MULTICAST PROTOCOLS: ODMRP, PUMA AND OBAMP*

Reference [38] have done simulation based analysis of the three multicast protocols namely On-Demand Multicast Routing Protocol (ODMRP), Protocol for Unified Multicasting through Announcements (PUMA), and Overlay Borùvka based Ad hoc Multicast Protocol (OBAMP). Packet Delivery Ratio, End-to-End Delay, and Total Bytes Send per Bytes Delivered were used as performance metrics as functions of Transmission Range, Mobility, and the group size. The environment chosen for evaluation of the protocol was ns-2.29. Data streams used were those of CBR. Results have proven that channel access of PUMA is more efficient to other two protocols. It was also observable that nearly constant end to end delay even with multiple senders.

*D. A Performance Comparison of On-Demand Multicast Routing Protocols for Ad Hoc Networks*

Reference [20] evaluated three reactive multicast routing protocols i.e. ADMR, MAODV, and ODMRP. The work focused on the impacts of changing number of participant sending and receiving nodes, changing number of total nodes, and the sending pattern. Packet Delivery Ratio, Control Packet Overhead, Normalized Packet Overhead, and End-to-End Delay were used as performance metrics. For mobility of nodes random waypoint mobility model was adopted. Nodes in the network move at random speed and to random destinations. Finally results are conducted from different perspectives i.e. by changing the quantity of multicast receivers, by changing the quantity of multicast sources, by changing the size of the network, by conferencing and so on. Results proved that all the protocols were performing well by means of packet delivery ratio, but ADMR proved to be better in terms of packet overhead by scaling its overhead strategy to satisfy communication needs.

*E. Multicasting in Ad-Hoc Networks: Comparing MAODV and ODMRP*

In their work [27] have evaluated and compared the performances of two of ad hoc multicast routing protocols. They have made the simulation based comparison by using

Network Simulator (NS2). Traffic used was constant bit rate (CBR). The performance metrics used were: Packet Delivery Ratio, Number of data packets sent per data packet received, Number of control packets transmitted per data packet received, and Number of control packets and data packets transmitted per data packet received. Each multicast group contained different number of nodes. The study concludes that ODMRP performed better to MAODV in dynamic environment.

## III. METHODOLOGY

### A. Approach

24 different scenarios are simulated, 12 for each protocol, with varying number of multicast listeners, and varying number of multicast sessions per node, as described below:
Total Simulation Time: 600 Seconds
Total Number of Vehicles: 100
Total Area: 10000 m x 1000 m
Number of Multicast Groups: 1
Mobility Model: Manhattan Mobility (Straight Highway)
Mobility of Vehicles: 80 Km/Hour – 110 Km/Hour
Transmission range of each vehicle: 1000 m

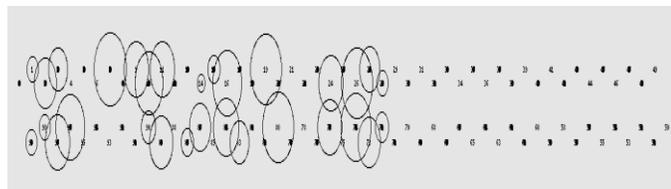

**Figure – I: A typical simulation in NS2**.

Scenario 1:
   Number of Simultaneous Listeners: 10
   Number of Sessions per Node: 5

Scenario 2:
   Number of Simultaneous Listeners: 10
   Number of Sessions per Node: 10

Scenario 3:
   Number of Simultaneous Listeners: 10
   Number of Sessions per Node: 20

Scenario 4:
   Number of Simultaneous Listeners: 20
   Number of Sessions per Node: 5

Scenario 5:
   Number of Simultaneous Listeners: 20
   Number of Sessions per Node: 10

Scenario 6:
   Number of Simultaneous Listeners: 20
   Number of Sessions per Node: 20

Scenario 7:
   Number of Simultaneous Listeners: 40
   Number of Sessions per Node: 5

Scenario 8:
   Number of Simultaneous Listeners: 40
   Number of Sessions per Node: 10

Scenario 9:
   Number of Simultaneous Listeners: 40
   Number of Sessions per Node: 20

Scenario 10:
   Number of Simultaneous Listeners: 60
   Number of Sessions per Node: 5

Scenario 11:
   Number of Simultaneous Listeners: 60
   Number of Sessions per Node: 10

Scenario 12:
   Number of Simultaneous Listeners: 60
   Number of Sessions per Node: 20

The reason behind varying number of simultaneous listeners is to testify the performances of protocols with varying extents of packet distribution. Whereas varying number of session per node causes reconfiguration of tree/mesh, so that we can examine whether the protocol remains consistent when its packet distribution structure is reconfigured frequently. Table - III summarizes the scenarios:

**Table – III: Simulations Summary**

| Routing Protocol | PUMA, MAODV |
|---|---|
| Simulator | NS2 |
| Total Vehicles | 100 |
| Mobility Model | Manhattan (Straight Highway) |
| Simulation Time | 600 Seconds |
| Simulation Area | 10,000 m X 1000 m |
| Simultaneous Listeners | 10, 20, 40, 60 |
| Listening Sessions per Node | 5, 10, 20 |
| Transmission Range | 1000 m |
| Average Data Packet Size | 512 bytes |
| Traffic Agent | UDP |
| Data Traffic | Variable Bit Rate (VBR) |

### B. Quality of Service Parameters

The quality metrics for evaluating the performance of the protocols in this work are the ones described in the following, the results are drawn from Trace (TR) files generated after simulating the scenarios.

*i. Packet Delivery Ratio (PDR):*

The packet delivery ratio is calculated as total number of packets received at destination per total number of packets supposed to be received. Mathematically,

PDR = (Received Packets)/(Packets_sould_be_received)

Where Packets_should_be_received is the count of packets those should be received according to number of listeners.

*ii. Average End-to-End Delay*

Average end-to-end delay is calculated by dividing to total E2E delay by total number of sent packets. Where total E2E delay is sum of delays incurred by each packet. The delay is measured in seconds. Mathematically,

Average EED = (Total EED)/(No of Sent Packets)

*iii. Throughput*

As the throughput measure is amount of data sent successfully per unit of time, we calculated total amount of sent data and divided it by total transmission time. Total transmission time is the time spanning from sending first data packet to sending last data packet. Mathematically:

Throughput = (Total amount of sent data)/(Total Transmission Time)

This formula calculate the amount of data sent in Bytes/Second. In order to calculate the throughout in Kbps, we calculated throughput as:

Throughput = (Total amount of sent data)/(Total Transmission Time)×(8/1024)

*iv. Normalized Routing Load (NRL)*

The NRL is the measure of control overhead a protocol incurs, thus it is calculated by number of control packets per data packet number data packets received divided:

NRL = (Number of Control Packets Sent) / (Number of Data Packets Received)

## C. Mobility Model

The Manhattan mobility model [16] from BonnMotion [6] is adapted to simulate vehicles movement on a highway. With this model, a node (vehicle) moves either horizontally or vertically, on x-axis or y-axis, that is quite similar to vehicles movement on the road. The vehicles move on predefined path that is an area of 10 Km x 1 Km with a random mobility speed between 80 Km/hour to 110 Km/hour. Area is divided into two portions, one for upstream and one for downstream of vehicles, on each portion two streams of vehicles are simulated, thus creating a real world scenario of vehicle movement on a highway.

## IV. RESULTS

### A. Throughput

The measure of amount of data received successfully per unit of time was observed with varying number of simultaneous listeners, that is, the number of multicast group members at one instance of time, hence increasing the number of join/leave sessions per node. The behavior of each protocol was observed as below:

*i. 10 Simultaneous Listeners*

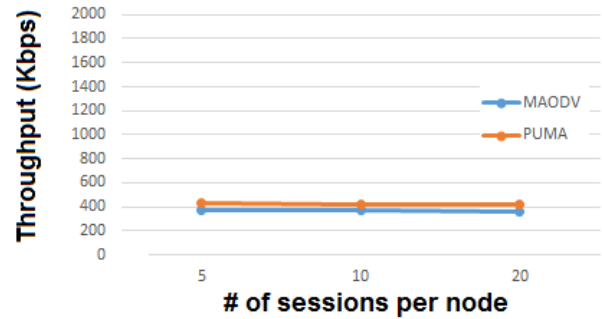

**Fig. – II : Throughput with 10 simultaneous listeners**

Figure II shows the throughput when number of simultaneous listeners are 10, observed throughput was 432.72 Kbps for PUMA, and 370.7 Kbps for MAODV, when each node was creating 5 sessions. Increasing the number of sessions per node, that is 10, 418.34 Kbps was observed throughput for PUMA, and 371.34 Kbps was observed for MAODV. Finally when each node waas creating 20 sessions, 422.21 Kbps was observed for PUMA, and 358.21 Kbps was observed for MAODV. Both the protocols are showing similar performances against increasining number of session per node.

*ii. 20 Simultaneous Listeners*

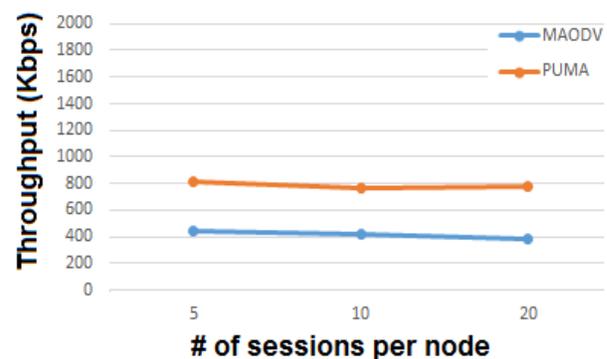

**Fig. – III : Throughput with 20 simultaneous listeners**

Figure III shows the throughput when number of simultaneous listeners are 20. With 20 simultaneous listeners, observed throughputs for PUMA were 808.92 Kbps, 763.86 Kbps, and 775.51 Kbps with 5, 10, and 20 sessions per node respectively. MAODV produced 438.92 Kbps, 419.86 Kbps, and 385.51 Kbps with 5, 10, and 20 sessions per node respectively. PUMA achieved a higher throughput for this scenario.

iii. 40 Simultaneous Listeners

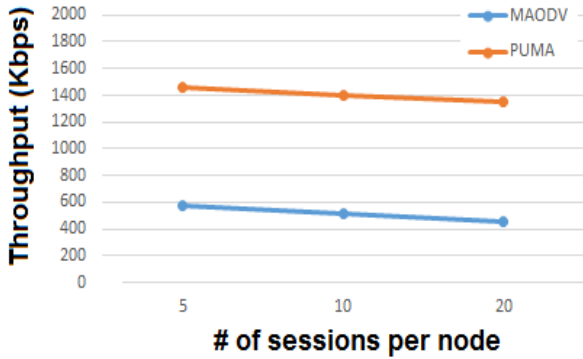

**Fig - IV: Throughput with 40 simultaneous listeners**

Figure - IV shows the throughput when simultaneous listener nodes are 40. When simultaneous listeners were 40, observed throughputs for PUMA were 1457.25 Kbps, 1400.73 Kbps, and 1354.63 Kbps with 5, 10, and 20 sessions per node respectively. MAODV, on the other hand produced 578.25 Kbps, 510.73 Kbps, and 454.87 Kbps with 5, 10, and 20 sessions per node respectively. PUMA, for this scenarios as well, achieved a higher throughput than MAODV.

iv. 60 Simultaneous Listeners

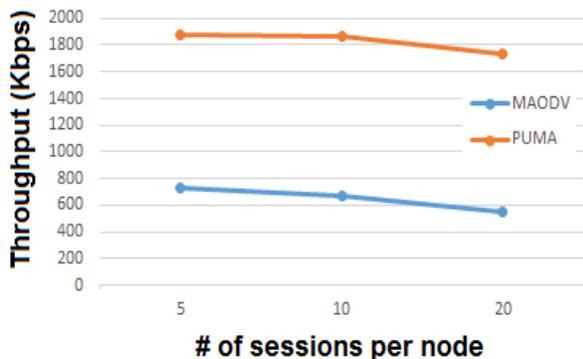

**Fig - V : Throughput with 60 simultaneous listeners**

Figure - V shows the throughput when number of simultaneous listeners are 60. Increasing simultaneous listeners to 60, PUMA produced these throughputs: 1878.84 Kbps with 5 sessions per node, 1868.34 Kbps with 10 sessions per node, and 1728.1 Kbps with 20 sessions per node. Observed throughputs for MAODV were 728.13 Kbps for 5 sessions per node, 668.19 Kbps for 10 sessions per node, and 552.81 Kbps for 20 sessions per node. PUMA outperformed MAODV in terms of throughput for this scenario.

B. *Normalized Routing Load*

Normalized Routing Load (NRL) describes the network overhead a protocol incurred for data transmission. Observed NRLs for each of PUMA and MAODV are described below.

i. 10 Simultaneous Listeners

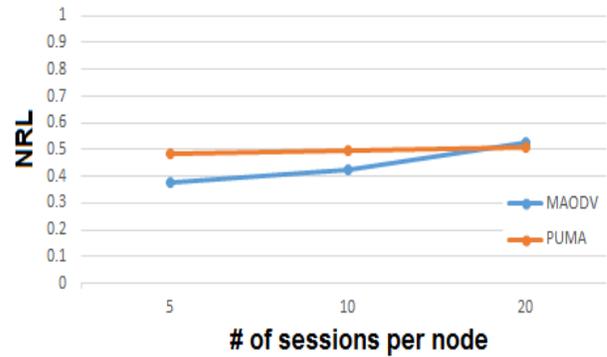

**Fig – VI : NRL with 10 simultaneous listeners**

Figure VI displays the normalized routing load with 10 simultaneous listener nodes, with 10 simultaneous listeners, observed NRLs for PUMA are: 0.485 with 5 sessions per node, 0.494 with 10 sessions per node, and 0.509 with 20 session per node. MAODV incurred these NRLs when simultaneous listeners are 10, 0.375 with 5 sessions per node, 0.424 with 10 sessions per node, and 0.527 with 20 sessions per node. MODV in this scenario showed inceasing NRL with increasing number of seesions per node.

ii. 20 Simultaneous Listeners

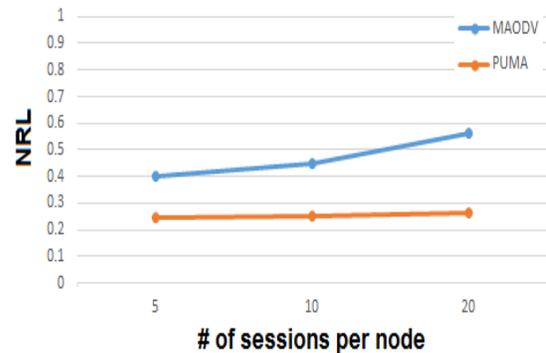

**Fig - VII: NRL with 20 simultaneous listeners**

Figure IV – 6 displays the normalized routing load with 20 simultaneous listener nodes, when increased the simultaneous listeners to 20, NRLs observed for PUMA were 0.245, 0.251, and 0.262 with 5, 10, and 20 sessions per node. With MAODV these are 0.401 with 5 sessions per node, 0.452 with 10 sessions per node, and 0.562 with 20 sessions per node. PUMA certainly dropped its NRL with increasing number of simultaneous listeners, and remained consistent against increase in number of sessions per node. MAODV on the other hand incurred more NRL with increase in number of simultaneous listeners, and also NRL increased for MAODV sith increasing number of sessions per node.

iii. 40 Simultaneous Listeners

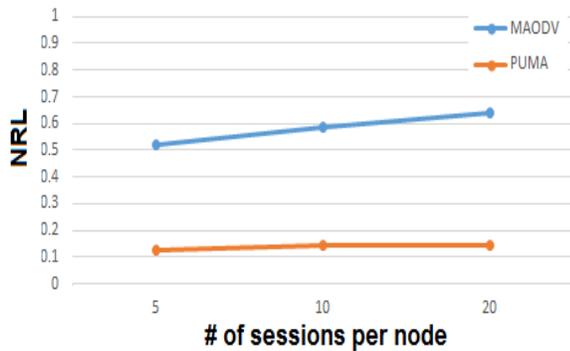

**Fig – VIII: NRL with 40 simultaneous listeners**

Figure – VIII displays the normalized routing load with 40 simultaneous listener nodes, with 40 simultaneous listeners PUMA incurred these NRLs: 0.128 with 5 sessions per node, 0.142 with 10 sessions per node, and 0.147 with 40 sessions per node. MAODV incurred these NRLs: 0.521 with 5 sessions per node, 0.587 with 10 sessions per node, and 0.641 with 20 sessions per node. Increasing number of simultaneous listeners did not impact NRL for PUMA, but MAODV continously incurred more NRL.

iv. 60 Simultaneous Listeners

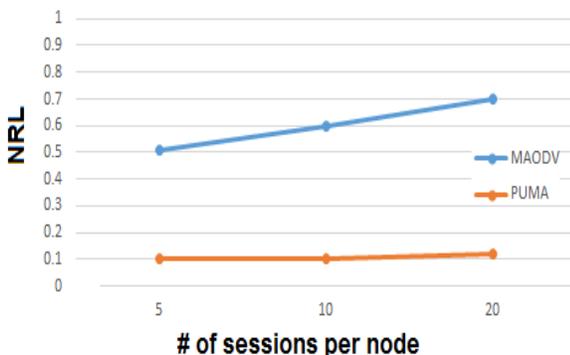

**Fig – IX: NRL with 60 simultaneous listeners**

Figure - IX displays the normalized routing load with 60 simultaneous listener nodes. With 60 simultaneous listeners NRLs for PUMA are: 0.102 with 5 sessions per node, 0.101 with 10 sessions per node, and 0.118 with 40 sessions per node. MAODV incurred these NRLs: 0.511 with 5 sessions per node, 0.601 with 10 sessions per node, and 0.7 with 20 sessions per node. PUMA achieved much better NRL for this scenario than MAODV which caused increasing NRL for this scenario as well.

C. Packet Delivery Ratio

For each of two protocols, observed fraction of received data packets to those which were sent is summarized below.

i. 10 Simultaneous Listeners

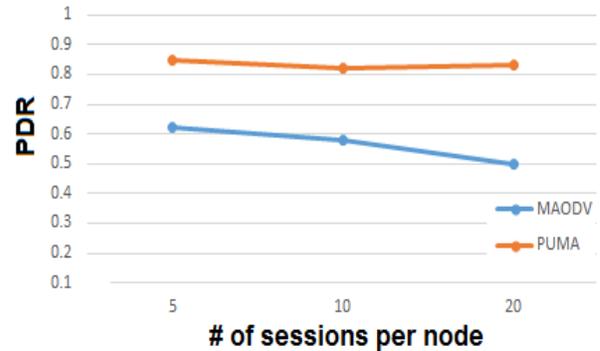

**Fig – X: PDR with 10 simultaneous listeners**

Figure – X shows the resulting Packet Delivery Ratio for both the routing protocols when simultaneous listeners are 10, with 10 simultaneous listeners, PUMA's PDRs remain 0.85 with 5 sessions per node, 0.82 with 10 sessions per node, and 0.83 with 20 sessions per node. MAODV's PDRs were 0.62 with 5 sessions per node, 0.58 with 10 sessions per node, and 0.50 with 20 sessions per node. PUMA achieved a higher packet delivery ratio than MAODV in this scenario.

ii. 20 Simultaneous Listeners

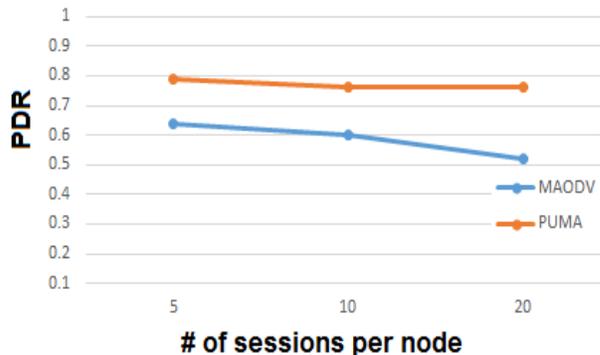

**Fig – XI: PDR with 20 simultaneous listeners**

Figure – XI shows the resulting Packet Delivery Ratio for both the routing protocols when simultaneous listeners are 20. When simultaneous listeners were 20, PUMA's PDR was 0.79 with 5 sessions per node, 0.76 with 10 sessions per node, and 0.76 with 20 sessions per node. MAODV's PDRs were 0.64 with 5 sessions per node, 0.60 with 10 sessions per node, and 0.52 with 20 sessions per node. PUMA achieved consistent PDR with increasing number of simultaneous listeners and increasing number of sessions per node. When simultaneous listeners are increased, MAODV drops down its PDR. Furthermore, MAODV's PDR dropped with increasing number of sessions per node.

iii. 40 Simultaneous Listeners

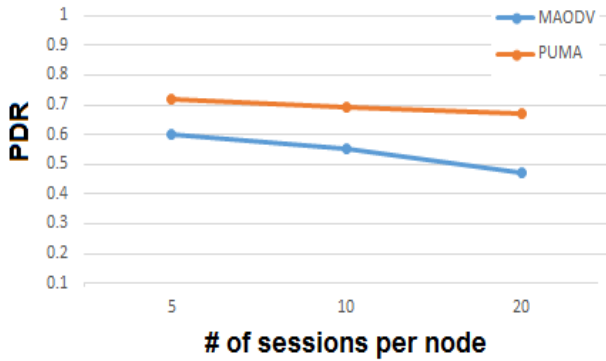

**Fig - XII: PDR with 40 simultaneous listeners**

Figure - XII shows the resulting Packet Delivery Ratio for both the routing protocols when simultaneous listeners are 40. With 40 simultaneous listeners PDRs for PUMA were 0.72, 0.69, and 0.67, and those of MAODV were 0.60, 0.55, and 0.47, with 5, 10, and 20 sessions per node respectively. Although, in this scenario PUMA slightly dropped down it PDR but it was still better than that of MAODV's. Furthermore, increasing number of sessions per node impacted more on MAODV's PDR than PUMA's PDR.

iv. 60 Simultaneous Listeners

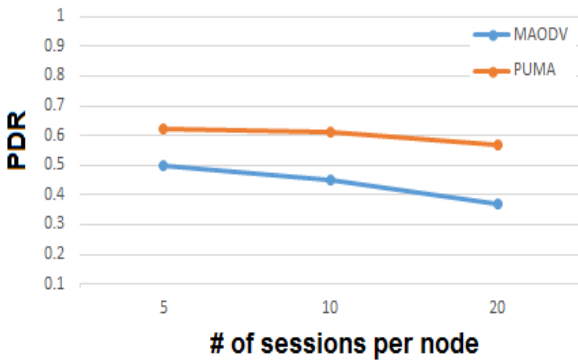

**Fig – XIII: PDR with 60 simultaneous listeners**

Figure - XIII shows the resulting Packet Delivery Ratio for both the routing protocols when simultaneous listeners are 60. With 60 simultaneous listeners PDRs for PUMA were: 0.62, 0.61, and 0.57, and those of MAODV were 0.50, 0.45, and 0.37, with 5, 10, and 20 sessions per node respectively. PUMA achieved better PDR than MAODV in this scenario as well.

D. *Average End-to-End Delay*

The average time taken by each protocol to transmit the data is described below, where this measure is take in seconds:

i. 10 Simultaneous Listeners

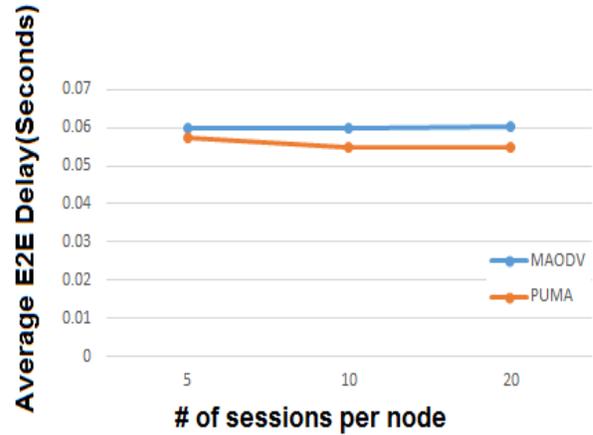

**Fig– XIV : Avg. E2E Delay with 10 simultaneous listeners**

Figure - XIV visualizes the Average End-to-End delay that each routing protocol incurred to route the data when simultaneous listeners are 10. With 10 simultaneous listeners PUMA showed these Avg. End-to-End Delays: 0.0575227 seconds (57.52 milliseconds) for 5 sessions per node, 0.054829 seconds (54.829 milliseconds) for 10 sessions per node, and 0.0547707 seconds (54.77 milliseconds) for 20 sessions per node. MAODV's Avg. E2E Delays were: 00.060 seconds (60 milliseconds) for 5 sessions per node, 0.060 seconds (60 milliseconds) for 10 sessions per node, and 0.06014 seconds (60.1 milliseconds) for 20 sessions per node. MAODV taken slightly more time to route the packet than PUMA in this scenario.

ii. 20 Simultaneous Listeners: 20

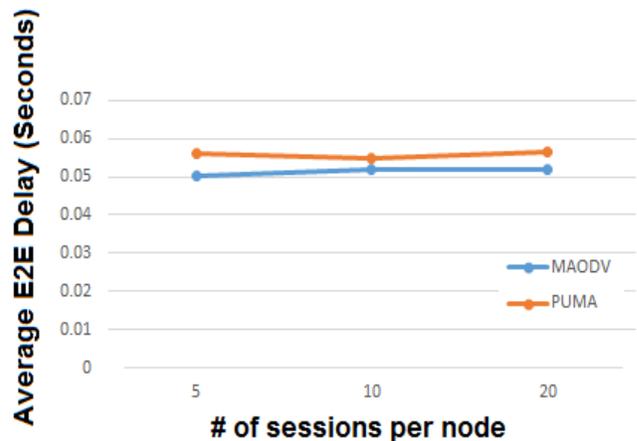

**Fig - XV: Avg. E2E Delay with 20 simultaneous listeners**

Figure - XV visualizes the Average End-to-End delay that each routing protocol incurred to route the data when simultaneous listeners are 20. When simultaneous listeners are 20, these are the Avg. End-to-End Delays of PUMA: 0.05589 seconds (55.89 milliseconds) for 5 sessions per node,

0.0547628 seconds (54.762 milliseconds) for 10 sessions per node, and 0.0565969 seconds (56.596 milliseconds) for 20 sessions per node. MAODV's Avg. E2E Delays were: 0.05012 seconds (50.12 milliseconds) for 5 sessions per node, 0.05176 seconds (51.76 milliseconds) for 10 sessions per node, and 0.05191 seconds (51.91 milliseconds) for 20 sessions per node. MAODV certainly dropped down it Avg. E2E delay with increase in number of simultaneous listeners. PUMA on the other hand remained consistent with increase in simultaneous listeners.

iii. 40 Simultaneous Listeners

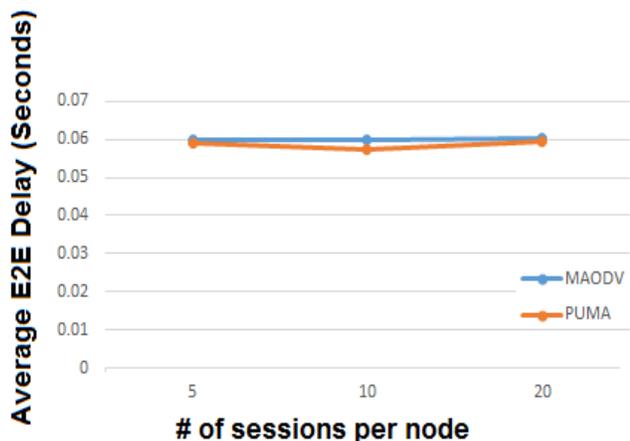

**Fig - XVI: Avg. E2E Delay with 40 Simultaneous Listeners**

Figure - XVI visualizes the Average End-to-End delay that each routing protocol incurred to route the data when simultaneous listeners are 40. Upon increasing simultaneous listeners to 40, the protocols showed these Avg. E2E Delays, PUMA: 0.058824 seconds (58.82 milliseconds) for 5 sessions per node, 0.0574805 seconds (57.48 milliseconds) for 10 sessions per node, and 0.059543 seconds (59.543 milliseconds) for 20 sessions per node. MAODV's Avg. E2E Delays were: 0.0598 seconds (59.8 milliseconds) for 5 sessions per node, 0.059801 seconds (59.8 milliseconds) for 10 sessions per node, and 0.0601 seconds (60.1 milliseconds) for 20 sessions per node. Both the protocols showed similar performances against Avg. E2E Delay in this scenario.

iv. 60 Simultaneous Listeners

Figure - XVII visualizes the Average End-to-End delay that each routing protocol incurred to route the data when simultaneous listeners are 60. Finally, with 60 simultaneous listeners, the protocols showed these Avg. E2E Delays, PUMA: 0.609066 seconds (60.90 milliseconds) for 5 sessions per node, 0.059735 seconds (59.73 milliseconds) for 10 sessions per node, and 0.059222 seconds (59.22 milliseconds) for 20 sessions per node. MAODV's Avg. E2E Delays were: 0.60 seconds (60 milliseconds) for 5 sessions per node, 0.05548 seconds (55.48 milliseconds) for 10 sessions per node, and 0.059222 seconds (59.22 milliseconds) for 20 sessions per node. MAODV at one instance (when number of sessions per node are 10) taken less time than PUMA, otherwise, both the protocols showed similar performance against Avg E2E Delay in this scenario.

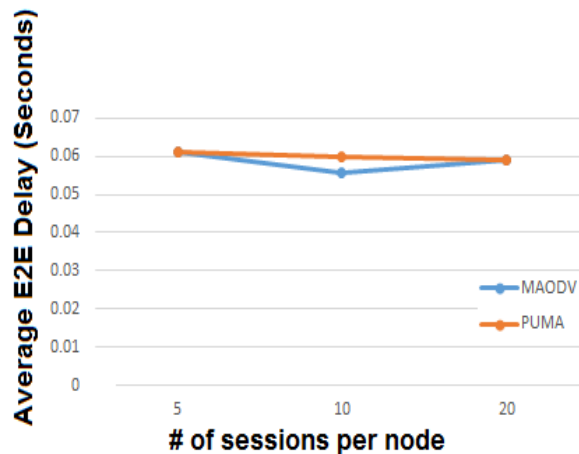

**Fig - XVII: Avg. E2E Delay with 60 Simultaneous Listeners**

## V. DISCUSSION

### A. Throughput

In different scenarios and, it can clearly be observed that PUMA outperforms the MAODV when comparing the protocols against throughput. Increasing the number of simultaneous listeners boosted the throughput achieved with PUMA, and slightly increased throughput with MAODV. Furthermore, increasing the number of join/leaves session per node did not impact much on throughput for both of the protocols.

### B. Normalized Routing Load

Be Initially with less number of simultaneous listeners, those are 10, PUMA incurred slightly higher NRL, but with increasing number of simultaneous listeners PUMA decreased down its NRL notably, because when the nodes which were simply forwarding the data become group members, they started receiving the data as well thus same controls packets are used to receive and forward the data. Redundant paths proved another better approach for PUMA as nodes were frequently joining/leaving the groups, there was little overhead of establishing new paths. MAODV on the other hand, increased its normalized routing load, this is because, every time listeners are increased, single distribution paths from sender to receiver are maintained, when nodes are leaving/joining the multicast group, more overhead is incurred to establish a path from source node to the destination node, furthermore, trees are the packet distribution structure those are more sensitive to topological changes.

### C. Packet Delivery Ratio

PUMA achieved a higher packet delivery ratio in all the scenarios than MAODV. The reason behind this is that PUMA

has redundant paths to deliver the data, if one of the path is lost, an alternative path may be followed to deliver the packet, MAODV on the other side has single path to each of receivers, losing that path causes loss of packets, and that path is needed to be re-established from the source.

*D. Avg. End-to-End Delay*

When compared against average end-to-end delay, both the protocols showed similar performances.

## VI. CONCLUSION

Among other developing technologies, video streaming can be taken as foreseeing trend in the area of VANET. This needs identification of a routing protocol that can efficiently support video transmission in VANETs. This study focuses on simulation based analysis of two multicast protocols (MAODV and PUMA) designed for ad hoc networks, using the video traffic. Different highway scenarios are taken to test the performance of these protocols individually under a simulated highway environment. Manhattan mobility model, that is similar to traffic movement on roads, is taken to create mobility of nodes, so that the created scenarios will be as much as realistic. The performances of the protocol are analyzed and compared on the basis of four QoS parameters (PDR, E2E Delay, Throughput, and the NRL).

PUMA proved better simulation results across all the scenarios with higher throughput and normalized routing load, and lesser normalized routing load. MAODV on the other hand remained comparatively better against E2E Delay.